\documentclass[pra,onecolumn,showkeys]{revtex4}%
\usepackage{amsfonts}
\usepackage{amsmath}
\usepackage{amssymb}
\usepackage{graphicx}
\usepackage{hyperref}%
\providecommand{\U}[1]{\protect\rule{.1in}{.1in}}
\newtheorem{theorem}{Theorem}

\newtheorem{remark}[theorem]{Remark}

\begin{document}
\title{Entropic uncertainty and measurement reversibility}
\author{Mario Berta}
\affiliation{Institute for Quantum Information and Matter, California Institute of
Technology, Pasadena, California 91125, USA}
\author{Stephanie Wehner}
\affiliation{QuTech, Delft University of Technology, Lorentzweg 1, 2628 CJ Delft, Netherlands}
\author{Mark M.~Wilde}
\affiliation{Hearne Institute for Theoretical Physics, Department of Physics and Astronomy,}
\affiliation{Center for Computation and Technology, Louisiana State University, Baton
Rouge, Louisiana 70803, USA}
\keywords{entropic uncertainty relations, measurement reversibility, monotonicity of
relative entropy}
\pacs{}

\begin{abstract}
The entropic uncertainty relation with quantum side information (EUR-QSI) from
[Berta \textit{et al}., Nat.~Phys.~\textbf{6}, 659 (2010)] is a unifying
principle relating two distinctive features of quantum mechanics:\ quantum
uncertainty due to measurement incompatibility, and entanglement. In these
relations, quantum uncertainty takes the form of preparation uncertainty where
\emph{one of two} incompatible measurements is applied. In particular, the
\textquotedblleft uncertainty witness\textquotedblright\ lower bound in the
EUR-QSI is not a function of a post-measurement state. An insightful proof of
the EUR-QSI from [Coles \textit{et al}., Phys.~Rev.~Lett.~\textbf{108}, 210405
(2012)]\ makes use of a fundamental mathematical consequence of the postulates
of quantum mechanics known as the non-increase of quantum relative entropy
under quantum channels. Here, we exploit this perspective to establish a
tightening of the EUR-QSI which adds a new state-dependent term in the lower
bound, related to how well one can reverse the action of a quantum
measurement. As such, this new term is a direct function of the
post-measurement state and can be thought of as quantifying how much
disturbance a given measurement causes. Our result thus quantitatively unifies
this feature of quantum mechanics with the others mentioned above. We have
experimentally tested our theoretical predictions on the IBM\ \textit{Quantum
Experience} and find reasonable agreement between our predictions and
experimental outcomes.

\end{abstract}
\volumeyear{ }
\volumenumber{ }
\issuenumber{ }
\eid{ }
\date{\today}
\startpage{1}
\endpage{10}
\maketitle

\section{Introduction}

The uncertainty principle is one of the cornerstones of modern physics,
providing a striking separation between classical and quantum mechanics
\cite{H27}. It is routinely used to reason about the behavior of quantum
systems, and in recent years, an information-theoretic refinement of it that
incorporates quantum side information has been helpful for witnessing
entanglement and in establishing the security of quantum key distribution
\cite{BCCRR10}. This latter refinement, known as the entropic uncertainty
relation with quantum side information (EUR-QSI), is the culmination of a
sequence of works spanning many decades
\cite{R29,hirschman57,bial1,beckner75,D83,K87,MU88,KP02,christandl05,RB09}%
\ and is the one on which we focus here.

\textbf{Tripartite uncertainty relations.} There are two variations of the
EUR-QSI \cite{BCCRR10}, one for tripartite and one for bipartite scenarios.
For the first, let $\rho_{ABE}$ denote a tripartite quantum state shared
between Alice, Bob, and Eve, and let $\mathbb{X}\equiv\{P_{A}^{x}\}$ and
$\mathbb{Z}=\{Q_{A}^{z}\}$ be projection-valued measures (PVMs) that can be
performed on Alice's system (note that considering PVMs implies statements for
the more general positive operator-valued measures, by invoking the Naimark
extension theorem \cite{Naimark40}). If Alice chooses to measure $\mathbb{X}$,
then the post-measurement state is as follows:
\begin{align}
\sigma_{XBE}  &  \equiv\sum_{x}|x\rangle\langle x|_{X}\otimes\sigma_{BE}%
^{x}\qquad\mathrm{where}\nonumber\\
\sigma_{BE}^{x}  &  \equiv\operatorname{Tr}_{A}\{(P_{A}^{x}\otimes I_{BE}%
)\rho_{ABE}\}. \label{eq:sig_X_state}%
\end{align}
Similarly, if Alice instead chooses to measure $\mathbb{Z}$, then the
post-measurement state is
\begin{align}
\omega_{ZBE}  &  \equiv\sum_{z}|z\rangle\langle z|_{Z}\otimes\omega_{BE}%
^{z}\qquad\mathrm{where}\nonumber\\
\omega_{BE}^{z}  &  \equiv\operatorname{Tr}_{A}\{(Q_{A}^{z}\otimes I_{BE}%
)\rho_{ABE}\}. \label{eq:om_Z_state}%
\end{align}
In the above, $\{|x\rangle_{X}\}_{x}$ and $\{|z\rangle_{Z}\}_{z}$ are
orthonormal bases that encode the classical outcome of the respective
measurements. The following tripartite EUR-QSI in \eqref{eq:EUR-QSI-1}
quantifies the trade-off between Bob's ability to predict the outcome of the
$\mathbb{X}$\ measurement with the help of his quantum system $B$ and Eve's
ability to predict the outcome of the $\mathbb{Z}$\ measurement with the help
of her system~$E$:
\begin{equation}
H(X|B)_{\sigma}+H(Z|E)_{\omega}\geq-\log c, \label{eq:EUR-QSI-1}%
\end{equation}
where here and throughout we take the logarithm to have base two. In the
above,
\begin{equation}
H(F|G)_{\tau}\equiv H(FG)_{\tau}-H(G)_{\tau}=H(\tau_{FG})-H(\tau_{G})
\end{equation}
denotes the conditional von Neumann entropy of a state $\tau_{FG}$, with
$H(\tau)\equiv-\operatorname{Tr}\{\tau\log\tau\}$, and\ the parameter $c$
captures the incompatibility of the $\mathbb{X}$ and $\mathbb{Z}$
measurements:
\begin{equation}
c\equiv\max_{x,z}\Vert P_{A}^{x}Q_{A}^{z}\Vert_{\infty}^{2}\in\lbrack0,1].
\label{eq:c-param}%
\end{equation}
The conditional entropy $H(F|G)_{\tau}$ is a measure of the uncertainty about
system $F$ from the perspective of someone who possesses system $G$, given
that the state of both systems is $\tau_{FG}$. The uncertainty relation in
\eqref{eq:EUR-QSI-1} thus says that if Bob can easily predict $X$ given $B$
(i.e., $H(X|B)$ is small) and the measurements are incompatible, then it is
difficult for Eve to predict $Z$ given $E$ (i.e., $H(Z|E)$ is large). As such,
\eqref{eq:c-param} at the same time captures measurement incompatibility and
the monogamy of entanglement~\cite{monogamyGames}. A variant of
\eqref{eq:EUR-QSI-1} in terms of the conditional min-entropy~\cite{TR11} can
be used to establish the security of quantum key distribution under particular
assumptions \cite{TLGR12,TL15}.

The EUR-QSI in \eqref{eq:EUR-QSI-1} can be summarized informally as a game
involving a few steps. To begin with, Alice, Bob, and Eve are given a state
$\rho_{ABE}$. Alice then flips a coin to decide whether to measure
$\mathbb{X}$ or $\mathbb{Z}$. If she gets heads, she measures $\mathbb{X}$ and
tells Bob that she did so. Bob then has to predict the outcome of her
$\mathbb{X}$\ measurement and can use his quantum system $B$ to help do so. If
Alice gets tails, she instead measures $\mathbb{Z}$ and tells Eve that she did
so. In this case, Eve has to predict the outcome of Alice's $\mathbb{Z}$
measurement and can use her quantum system $E$ as an aid. There is a trade-off
between their ability to predict correctly, which is captured by \eqref{eq:EUR-QSI-1}.

\textbf{Bipartite uncertainty relations.} We now recall the second variant of
the EUR-QSI from~\cite{BCCRR10}. Here we have a bipartite state $\rho_{AB}%
$\ shared between Alice and Bob and again the measurements $\mathbb{X}$ and
$\mathbb{Z}$ mentioned above. Alice chooses to measure either $\mathbb{X}$ or
$\mathbb{Z}$, leading to the respective post-measurement states $\sigma_{XB}$
and $\omega_{ZB}$ defined from \eqref{eq:sig_X_state} and
\eqref{eq:om_Z_state} after taking a partial trace over the $E$ system. The
following EUR-QSI in \eqref{eq:EUR-QSI-2} quantifies the trade-off between
Bob's ability to predict the outcome of the $\mathbb{X}$ or $\mathbb{Z}$
measurement:
\begin{equation}
H( Z|B) _{\omega}+H( X|B) _{\sigma}\geq-\log c+H( A|B) _{\rho},
\label{eq:EUR-QSI-2}%
\end{equation}
where the incompatibility parameter $c$ is defined in \eqref{eq:c-param} and
the conditional entropy $H( A|B) _{\rho}$ is a signature of both the mixedness
and entanglement of the state $\rho_{AB}$. For \eqref{eq:EUR-QSI-2} to hold,
we require the technical condition that the $\mathbb{Z}$ measurement be a
rank-one measurement \cite{CYZ11} (however see also \cite{FL13,furrer13} for a
lifting of this condition). The EUR-QSI in \eqref{eq:EUR-QSI-2} finds
application in witnessing entanglement, as discussed in \cite{BCCRR10}.

The uncertainty relation in \eqref{eq:EUR-QSI-2} can also be summarized
informally as a game, similar to the one discussed above. Here, we have Alice
choose whether to measure $\mathbb{X}$ or $\mathbb{Z}$. If she measures
$\mathbb{X}$, she informs Bob that she did so, and it is his task to predict
the outcome of the $\mathbb{X}$ measurement. If she instead measures
$\mathbb{Z}$, she tells Bob, and he should predict the outcome of the
$\mathbb{Z}$ measurement. In both cases, Bob is allowed to use his quantum
system $B$ to help in predicting the outcome of Alice's measurement. Again
there is generally a trade-off between how well Bob can predict the outcome of
the $\mathbb{X}$ or $\mathbb{Z}$ measurement, which is quantified by
\eqref{eq:EUR-QSI-2}. The better that Bob can predict the outcome of either
measurement, the more entangled the state $\rho_{AB}$ is.

\section{Main result}

The main contribution of the present paper is to refine and tighten both of the
uncertainty relations in \eqref{eq:EUR-QSI-1} and \eqref{eq:EUR-QSI-2} by
employing a recent result from~\cite{Junge15} (see also \cite{W15,Sutter15,SBT16}).
This refinement adds a term involving measurement reversibility, next to the
original trade-offs in terms of measurement incompatibility and entanglement.
An insightful proof of the EUR-QSIs above makes use of an entropy inequality
known as the non-increase of quantum relative entropy \cite{Lindblad1975,U77}.
This entropy inequality is fundamental in quantum physics, providing
limitations on communication protocols \cite{W13} and thermodynamic processes
\cite{S12}. The main result of \cite{Junge15,W15,Sutter15,SBT16} offers a
strengthening of the non-increase of quantum relative entropy,
quantifying\ how well one can recover from the deleterious effects of a noisy
quantum channel. Here we apply the particular result from \cite{Junge15} to
establish a tightening of both uncertainty relations in \eqref{eq:EUR-QSI-1}
and \eqref{eq:EUR-QSI-2} with a term related to how well one can
\textquotedblleft reverse\textquotedblright\ an additional $\mathbb{X}$
measurement performed on Alice's system at the end of the uncertainty game, if
the outcome of the $\mathbb{X}$\ measurement and the $B$ system are available.
The upshot is an entropic uncertainty relation which incorporates measurement
reversibility in addition to quantum uncertainty due to measurement
incompatibility, and entanglement, thus unifying several genuinely quantum
features into a single uncertainty relation.

In particular, we establish the following refinements of \eqref{eq:EUR-QSI-1}
and \eqref{eq:EUR-QSI-2}:
\begin{align}
H(Z|E)_{\omega}+H(X|B)_{\sigma}  &  \geq-\log c-\log
f,\label{eq:EUR-QSI-1-new}\\
H(Z|B)_{\omega}+H(X|B)_{\sigma}  &  \geq-\log c-\log f+H(A|B)_{\rho},
\label{eq:EUR-QSI-2-new}%
\end{align}
where $c$ is defined in \eqref{eq:c-param},
\begin{equation}
f\equiv F(\rho_{AB},\mathcal{R}_{XB\rightarrow AB}(\sigma_{XB})),
\end{equation}
and in \eqref{eq:EUR-QSI-2-new} we need the projective $\mathbb{Z}$
measurement to be a rank-one measurement (i.e., $Q_{A}^{z}=|z\rangle\langle
z|$). In addition to the measurement incompatibility $c$, the term $f$
quantifies the disturbance caused by one of the measurements, in particular,
how reversible such a measurement is. $F(\rho_{1},\rho_{2})\equiv\Vert
\sqrt{\rho_{1}}\sqrt{\rho_{2}}\Vert_{1}^{2}$ denotes the quantum fidelity
between two density operators $\rho_{1}$ and $\rho_{2}$ \cite{U73}, and
$\mathcal{R}_{XB\rightarrow AB}$ is a \textit{recovery} quantum channel with
input systems $XB$ and output systems $AB$. Appendix~\ref{app:proof} details a
proof for \eqref{eq:EUR-QSI-1-new} and~\eqref{eq:EUR-QSI-2-new}. In Section~\ref{sec:examples}, we discuss several simple exemplary states and measurements to which \eqref{eq:EUR-QSI-2-new} applies, and in Section~\ref{sec:experiments}, we detail the results of several experimental tests of the theoretical predictions, finding reasonable agreement between the experimental results and our predictions.

In the case that the $\mathbb{Z}$ measurement has the form $\{Q_{A}%
^{z}=|z\rangle\langle z|_{A}\}_{z}$ for an orthonormal basis $\{|z\rangle
_{A}\}_{z}$, the action of the recovery quantum channel $\mathcal{R}%
_{XB\rightarrow AB}$ on an arbitrary state $\xi_{XB}$ is explicitly given as
follows (see Appendix~\ref{sec:recovery-map-explicit}\ for details):%
\begin{equation}
\mathcal{R}_{XB\rightarrow AB}(\xi_{XB})=\sum_{z,x,z^{\prime}}|z\rangle\langle
z|_{A}P_{A}^{x}|z^{\prime}\rangle\langle z^{\prime}|_{A}\otimes\mathcal{R}%
_{XB\rightarrow B}^{x,z,z^{\prime}}(\xi_{XB}), \label{eq:main:recovery1}%
\end{equation}
where%
\begin{equation}
\mathcal{R}_{XB\rightarrow B}^{x,z,z^{\prime}}(\xi_{XB})\equiv\int_{-\infty
}^{\infty}dt\ p(t)\left(  \omega_{B}^{z}\right)  ^{\frac{1-it}{2}}\left(
\theta_{B}^{x}\right)  ^{\frac{-1+it}{2}}\operatorname{Tr}_{X}\{|x\rangle
\langle x|_{X}(\xi_{XB})\}\left(  \theta_{B}^{x}\right)  ^{\frac{-1-it}{2}%
}(\omega_{B}^{z^{\prime}})^{\frac{1+it}{2}},
\end{equation}
with the probability density $p(t)\equiv\frac{\pi}{2}(\cosh(\pi t)+1)^{-1}$.
(Note that $\mathcal{R}_{XB\rightarrow B}^{x,z,z^{\prime}}$ is not a
channel---we are merely using this notation as a shorthand.) In the above,
$\theta_{XB}$ is the state resulting from Alice performing the $\mathbb{Z}%
$\ measurement, following with the $\mathbb{X}$\ measurement, and then
discarding the outcome of the $\mathbb{Z}$\ measurement:%
\begin{align}
\theta_{XB}  &  \equiv\sum_{x}|x\rangle\langle x|_{X}\otimes\theta_{B}%
^{x}\quad\mathrm{with}\nonumber\\
\theta_{B}^{x}  &  \equiv\sum_{z}\langle z|P_{A}^{x}|z\rangle\ \omega^{z}_{B}.
\label{eq:theta-state}%
\end{align}
For this case, $\omega^{z}_{B}$ from \eqref{eq:om_Z_state} reduces to
$\omega^{z}_{B} = (\langle z|_{A}\otimes I_{B})\rho_{AB}(|z\rangle_{A}\otimes
I_{B})$. As one can readily check by plugging into \eqref{eq:main:recovery1},
the recovery channel $\mathcal{R}$ has the property that it perfectly reverses
an $\mathbb{X}$\ measurement if it is performed after a $\mathbb{Z}%
$\ measurement:%
\begin{equation}
\mathcal{R}_{XB\rightarrow AB}(\theta_{XB})=\sum_{z}|z\rangle\langle
z|_{A}\otimes\omega^{z}_{B}. \label{eq:Z-X-recover-perfect}%
\end{equation}

The fidelity $F(\rho_{AB},\mathcal{R}_{XB\rightarrow AB}(\sigma_{XB}))$ thus
quantifies how much disturbance the $\mathbb{X}$\ measurement causes to the
original state $\rho_{AB}$ in terms of how well the recovery channel
$\mathcal{R}$ can reverse the process. We note that there is a trade-off
between reversing the $\mathbb{X}$ measurement whenever it is greatly
disturbing $\rho_{AB}$ and meeting the constraint in
\eqref{eq:Z-X-recover-perfect}. Since the quantum fidelity always takes a
value between zero and one, it is clear that \eqref{eq:EUR-QSI-1-new} and
\eqref{eq:EUR-QSI-2-new} represent a state-dependent tightening of
\eqref{eq:EUR-QSI-1} and \eqref{eq:EUR-QSI-2}, respectively.

\section{Interpretation}

It is interesting to note that just as the original relation in
\eqref{eq:EUR-QSI-2} could be used to witness entanglement, the new relation
can be used to witness both entanglement and recovery from measurement, as
will be illustrated using the examples below. That is, having low conditional
entropy for both measurement outcomes constitutes a recoverability witness,
when given information about the entanglement.

\begin{figure}[ptb]
\begin{center}
\includegraphics[
width=6.0456in
]{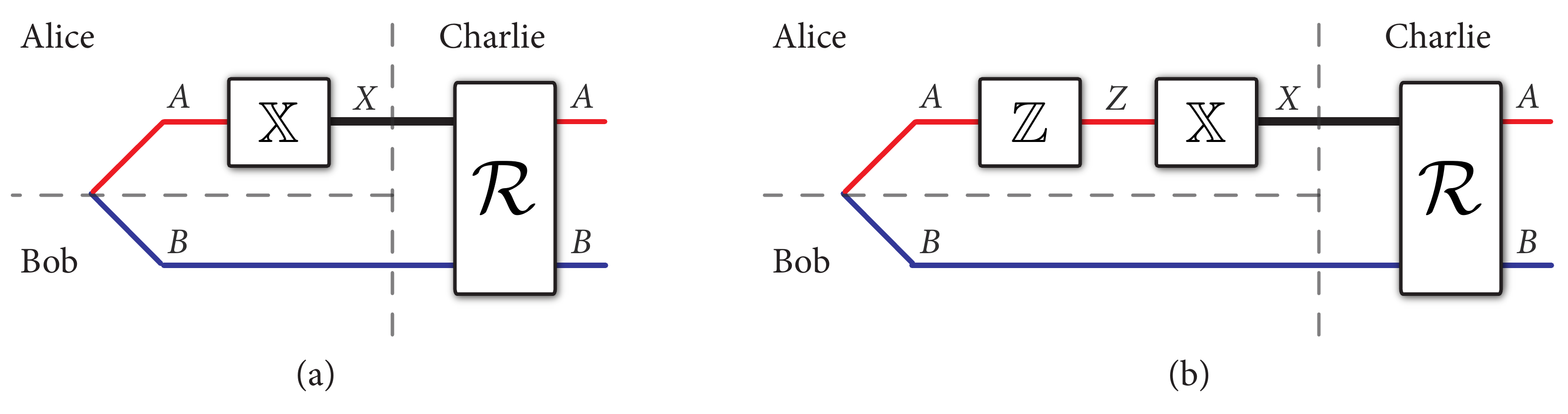}
\end{center}
\caption{\textbf{Measurement reversibility game.} How well can Charlie reverse
the action of the $\mathbb{X}$ measurement in either scenario (a) or (b)? The
quantities in \eqref{eq:EUR-QSI-1-new} and \eqref{eq:EUR-QSI-2-new} other than
$f$ constitute a ``recoverability witness,'' quantifying Charlie's ability to
do so.}%
\label{fig:meas-reverse-game}%
\end{figure}

We recalled above the established \textquotedblleft uncertainty
games\textquotedblright\ in order to build an intuition for
\eqref{eq:EUR-QSI-1} and \eqref{eq:EUR-QSI-2}. In order to further understand
the refinements in \eqref{eq:EUR-QSI-1-new} and \eqref{eq:EUR-QSI-2-new}, we
could imagine that after either game is completed, we involve another player
Charlie. Regardless of which measurement Alice performed in the original game,
she then performs an additional $\mathbb{X}$ measurement. Bob sends his
quantum system $B$ to Charlie, and Alice sends the classical outcome of the
final $\mathbb{X}$ measurement to Charlie. It is then Charlie's goal to
\textquotedblleft reverse\textquotedblright\ the $\mathbb{X}$ measurement in
either of the scenarios above, and his ability to do so is limited by the
uncertainty relations in \eqref{eq:EUR-QSI-1-new} and
\eqref{eq:EUR-QSI-2-new}. Figure~\ref{fig:meas-reverse-game} depicts this
game. In the case that (a) Alice performed an $\mathbb{X}$ measurement in the
original game, the state that Charlie has is $\sigma_{XB}$. In the case that
(b)\ Alice performed a $\mathbb{Z}$ measurement in the original game, then the
state that Charlie has is $\theta_{XB}$. Not knowing which state he has
received, Charlie can perform the recovery channel $\mathcal{R}$ and be
guaranteed to restore the state to%
\begin{equation}
\sum_{z}|z\rangle\langle z|\otimes\big(\langle z|\otimes I_{B}\big)\rho
_{AB}\big(|z\rangle\otimes I_{B}\big)
\end{equation}
in the case that (b)\ occurred, while having a performance limited by
\eqref{eq:EUR-QSI-1-new} or \eqref{eq:EUR-QSI-2-new} in the case that (a) occurred.

\section{Examples}

\label{sec:examples}

It is helpful to examine some examples in order to build an intuition for our
refinements of the EUR-QSIs. Here we focus on the bipartite EUR-QSI
in~\eqref{eq:EUR-QSI-2-new} and begin by evaluating it for some ``minimum
uncertainty states''~\cite{CYZ11} (see also~\cite{CCYZ12}). These are states
for which the original uncertainty relation in \eqref{eq:EUR-QSI-2} is already
tight, i.e., an equality. Later, we will consider the case of a representative
``maximum uncertainty state,'' that is, a state for which the original
uncertainty relation~\eqref{eq:EUR-QSI-2} is maximally non-tight. This last
example distinguishes our new contribution in~\eqref{eq:EUR-QSI-2-new} from
the previously established bound in~\eqref{eq:EUR-QSI-2}.

For all of the forthcoming examples, we take the $\mathbb{X}$ measurement to
be Pauli $\sigma_{X}$ and the $\mathbb{Z}$ measurement to be Pauli $\sigma
_{Z}$, which implies that $-\log c=1$. We define the \textquotedblleft
BB84\textquotedblright\ states $|0\rangle$, $|1\rangle$, $|+\rangle$, and
$|-\rangle$ from the following relations:%
\begin{equation}
\sigma_{Z}|0\rangle=|0\rangle,\qquad\sigma_{Z}|1\rangle=(-1)|1\rangle
,\qquad\sigma_{X}|+\rangle=|+\rangle,\qquad\sigma_{X}|-\rangle=(-1)|-\rangle.
\end{equation}
So this means that the $\mathbb{X}$ and $\mathbb{Z}$ measurements have the
following respective implementations as quantum channels acting on an
input~$\xi$:
\begin{align}
\xi &  \rightarrow\langle+|_{A}\xi|+\rangle_{A}|0\rangle\langle0|_{X}%
+\langle-|_{A}\xi|-\rangle_{A}|1\rangle\langle1|_{X},\\
\xi &  \rightarrow\langle0|_{A}\xi|0\rangle_{A}|0\rangle\langle0|_{Z}%
+\langle1|_{A}\xi|1\rangle_{A}|1\rangle\langle1|_{Z}.
\end{align}

\subsection{Minimum uncertainty states}

\subsubsection{$X$ eigenstate on system $A$}

First suppose that $\rho_{AB}=|+\rangle\langle+|_{A}\otimes\pi_{B}$, where
$\pi$ is the maximally mixed state. In this case, Bob's system $B$ is of no
use to help predict the outcome of a measurement on the $A$ system because the
systems are in a product state. Here we find by direct calculation that
$H(A|B)_{\rho}=0$, $H(X|B)_{\sigma}=0$, and $H(Z|B)_{\omega}=1$. By
\eqref{eq:EUR-QSI-2-new}, this then implies that there exists a recovery
channel $\mathcal{R}^{(1)}$\ such that \eqref{eq:Z-X-recover-perfect} is
satisfied and, given that $\sigma_{XB}=|0\rangle\langle0|_{X}\otimes\pi_{B}$,
we also have the perfect recovery
\begin{equation}
\mathcal{R}_{XB\rightarrow AB}^{(1)}\big(|0\rangle\langle0|_{X}\otimes\pi
_{B}\big)=|+\rangle\langle+|_{A}\otimes\pi_{B}.\label{eq:r1-recovery}%
\end{equation}
To determine the recovery channel $\mathcal{R}^{(1)}$, consider that%
\begin{equation}
\sum_{z}|z\rangle\langle z|_{Z}\otimes\omega_{B}^{z}=\pi_{Z}\otimes\pi
_{B},\qquad\sum_{x}|x\rangle\langle x|_{X}\otimes\theta_{B}^{x}=\pi_{X}%
\otimes\pi_{B},
\end{equation}
with the states on the left in each case defined in \eqref{eq:om_Z_state}\ and
\eqref{eq:theta-state}, respectively. Plugging into \eqref{eq:main:recovery1},
we find that the recovery channel in this case is given explicitly by%
\begin{equation}
\mathcal{R}_{XB\rightarrow AB}^{(1)}(\xi_{XB})=|+\rangle\langle+|_{A}%
\otimes\operatorname{Tr}_{X}\{|0\rangle\langle0|_{X}\xi_{XB}\}+|-\rangle
\langle-|_{A}\otimes\operatorname{Tr}_{X}\{|1\rangle\langle1|_{X}\xi
_{XB}\},\label{eq:r1-recovery-X-eigen}%
\end{equation}
so that we also see that%
\begin{equation}
\mathcal{R}_{XB\rightarrow AB}^{(1)}(\pi_{X}\otimes\pi_{B})=\pi_{A}\otimes
\pi_{B}.\label{eq:r1-recovery-theta-omega}%
\end{equation}

\subsubsection{$Z$ eigenstate on system $A$}

The situation in which $\rho_{AB}=|0\rangle\langle0|_{A}\otimes\pi_{B}$ is
similar in some regards, but the recovery channel is different---i.e., we have by direct calculation that
$H(A|B)_{\rho}=0$, $H(X|B)_{\sigma}=1$, and $H(Z|B)_{\omega}=0$, which implies
the existence of a different recovery channel $\mathcal{R}^{(2)}$ such that
\eqref{eq:Z-X-recover-perfect} is satisfied, and given that $\sigma_{XB}%
=\pi_{X}\otimes\pi_{B}$, we also have the perfect recovery
\begin{equation}
\mathcal{R}_{XB\rightarrow AB}^{(2)}\big(\pi_{X}\otimes\pi_{B}\big)=|0\rangle
\langle0|_{A}\otimes\pi_{B}.
\end{equation}
To determine the recovery channel $\mathcal{R}^{(2)}$, consider that%
\begin{equation}
\sum_{z}|z\rangle\langle z|_{Z}\otimes\omega_{B}^{z}=|0\rangle\langle
0|_{Z}\otimes\pi_{B},\qquad\sum_{x}|x\rangle\langle x|_{X}\otimes\theta
_{B}^{x}=\pi_{X}\otimes\pi_{B},
\end{equation}
with the states on the left in each case defined in \eqref{eq:om_Z_state}\ and
\eqref{eq:theta-state}, respectively. Plugging into \eqref{eq:main:recovery1},
we find that the recovery channel in this case is given explicitly by%
\begin{equation}
\mathcal{R}_{XB\rightarrow AB}^{(2)}(\xi_{XB})=|0\rangle\langle0|_{A}%
\otimes\operatorname{Tr}_{X}\{\xi_{XB}\}.
\end{equation}

\subsubsection{Maximally entangled state on systems $A$ and $B$}

\label{sec:max-entangled-example}Now suppose that $\rho_{AB}=|\Phi
\rangle\langle\Phi|_{AB}$ is the maximally entangled state, where
$|\Phi\rangle_{AB}\equiv(|00\rangle_{AB}+|11\rangle_{AB})/\sqrt{2}$. In this
case, we have that both $H(X|B)_{\sigma}=0$ and $H(Z|B)_{\omega}=0$, but the
conditional entropy is negative:\ $H(A|B)_{\rho}=-1$. So here again we find
the existence of a recovery channel $\mathcal{R}^{(3)}$ such that
\eqref{eq:Z-X-recover-perfect} is satisfied, and given that $\sigma
_{XB}=(|0+\rangle\langle0+|_{XB}+|1-\rangle\langle1-|_{XB})/2$, we also have
the perfect recovery%
\begin{equation}
\mathcal{R}_{XB\rightarrow AB}^{(3)}\Big(\big(|0+\rangle\langle0+|_{XB}%
+|1-\rangle\langle1-|_{XB}\big)/2\Big)=|\Phi\rangle\langle\Phi|_{AB}%
.\label{eq:r3-recovery}%
\end{equation}
To determine the recovery channel $\mathcal{R}^{(3)}$, consider that%
\begin{align}
\sum_{z}|z\rangle\langle z|_{Z}\otimes\omega_{B}^{z} &  =\frac{1}{2}\left(
|0\rangle\langle0|_{Z}\otimes|0\rangle\langle0|_{B}+|1\rangle\langle
1|_{Z}\otimes|1\rangle\langle1|_{B}\right)  ,\\
\sum_{x}|x\rangle\langle x|_{X}\otimes\theta_{B}^{x} &  =\pi_{X}\otimes\pi
_{B},
\end{align}
with the states on the left in each case defined in \eqref{eq:om_Z_state}\ and
\eqref{eq:theta-state}, respectively. Plugging into \eqref{eq:main:recovery1},
we find that the recovery channel in this case is given explicitly by%
\begin{equation}
\mathcal{R}_{XB\rightarrow AB}^{(3)}(\xi_{XB})=\sum_{z,z^{\prime},x\in\left\{
0,1\right\}  }\left(  -1\right)  ^{x(z+z^{\prime})}|z\rangle\langle z^{\prime
}|_{A}\otimes|z\rangle\langle z^{\prime}|_{B}\operatorname{Tr}\{|x\rangle
\langle x|_{X}\otimes|z^{\prime}\rangle\langle z|_{B}\xi_{XB}%
\},\label{eq:max-ent-state-recovery-op}%
\end{equation}
i.e., with the following Kraus operators:%
\begin{equation}
\left\{  \sum_{z}\left(  -1\right)  ^{xz}\left(  |z\rangle_{A}\otimes
|z\rangle_{B}\right)  \left(  \langle x|_{X}\otimes\langle z|_{B}\right)
\right\}  _{x}.
\end{equation}
These Kraus operators give the recovery map $\mathcal{R}_{XB\rightarrow
AB}^{(3)}$ the interpretation of 1)\ measuring the $X$ register and
2)\ coherently copying the contents of the $B$ register to the $A$ register
along with an appropriate relative phase. It can be implemented by performing
a controlled-NOT gate from $B$ to $A$, followed by a controlled-phase gate on
$X$ and $B$ and a partial trace over system $X$.

\begin{remark}
All of the examples mentioned above involve a perfect recovery or a perfect
reversal of the $\mathbb{X}$ measurement. This is due to the fact that the
bound in \eqref{eq:EUR-QSI-2} is saturated for these examples. However, the
refined inequality in \eqref{eq:EUR-QSI-2-new} allows to generalize these
situations to the approximate case, in which $\rho_{AB}$ is nearly
indistinguishable from the states given above. It is then the case that the
equalities in \eqref{eq:r1-recovery}--\eqref{eq:r3-recovery} become
approximate equalities, with a precise characterization given by~\eqref{eq:EUR-QSI-2-new}.
\end{remark}

\subsection{Maximum uncertainty states}

We now investigate the extreme opposite situation, when the bound in
\eqref{eq:EUR-QSI-2} is far from being saturated but its refinement in
\eqref{eq:EUR-QSI-2-new} is saturated. Let $\rho_{AB}=|+_{Y}\rangle
\langle+_{Y}|_{A}\otimes\pi_{B}$, where $|+_{Y}\rangle$ is defined from the
relation $\sigma_{Y}|+_{Y}\rangle=|+_{Y}\rangle$. In this case, we find that
both $H(X|B)_{\sigma}=1$ and $H(Z|B)_{\omega}=1$. Thus, we could say that
$\rho_{AB}$ is a \textquotedblleft maximum uncertainty state\textquotedblright%
\ because the sum $H(X|B)_{\sigma}+H(Z|B)_{\omega}$ is equal to two bits and
cannot be any larger than this amount. We also find that $H(A|B)_{\rho}=0$,
implying that \eqref{eq:EUR-QSI-2} is one bit away from being saturated. Now
consider that $\sigma_{XB}=\theta_{XB}=\pi_{X}\otimes\pi_{B}$ and $\omega
_{ZB}=\pi_{Z}\otimes\pi_{B}$, and thus one can explicitly calculate the
recovery channel $\mathcal{R}^{(4)}$ from~\eqref{eq:main:recovery1} to take
the form:
\begin{equation}
\mathcal{R}_{XB\rightarrow AB}^{(4)}(\xi_{XB})\equiv|+\rangle\langle
+|_{A}\otimes\operatorname{Tr}_{X}\{|0\rangle\langle0|_{X}\xi_{XB}%
\}+|-\rangle\langle-|_{A}\otimes\operatorname{Tr}_{X}\{|1\rangle\langle
1|_{X}\xi_{XB}\}.
\end{equation}
Note that the recovery channel $\mathcal{R}_{XB\rightarrow AB}^{(4)}$ is the
same as $\mathcal{R}_{XB\rightarrow AB}^{(1)}$ in \eqref{eq:r1-recovery-X-eigen}.

This implies that%
\begin{align}
\mathcal{R}_{XB\rightarrow AB}^{(4)}(\sigma_{XB})  & =\pi_{A}\otimes\pi
_{B},\label{eq:r4-recovery-sigma-rho}\\
\mathcal{R}_{XB\rightarrow AB}^{(4)}(\theta_{XB})  & =\pi_{A}\otimes\pi
_{B},\label{eq:r4-recovery-theta-omega}%
\end{align}
and in turn that
\begin{equation}
-\log F(\rho_{AB},\mathcal{R}_{XB\rightarrow AB}^{(4)}(\theta_{XB}))=1.
\end{equation}
Thus the inequality in \eqref{eq:EUR-QSI-2-new} is saturated for this example.
The key element is that there is one bit of uncertainty when measuring a $Y$
eigenstate with respect to either the $X$ or $Z$ basis. At the same time, the
$Y$ eigenstate is pure, so that its entropy is zero. This leaves a bit of
uncertainty available and for which \eqref{eq:EUR-QSI-2} does not account, but
which we have now interpreted in terms of how well one can reverse the
$\mathbb{X}$ measurement, using the refined bound in \eqref{eq:EUR-QSI-2-new}.
One could imagine generalizing the idea of this example to higher dimensions
in order to find more maximum uncertainty examples of this sort.

\section{Experiments}

\label{sec:experiments}

\begin{figure}[ptb]
\begin{center}
\includegraphics[
width=4.9786in
]
{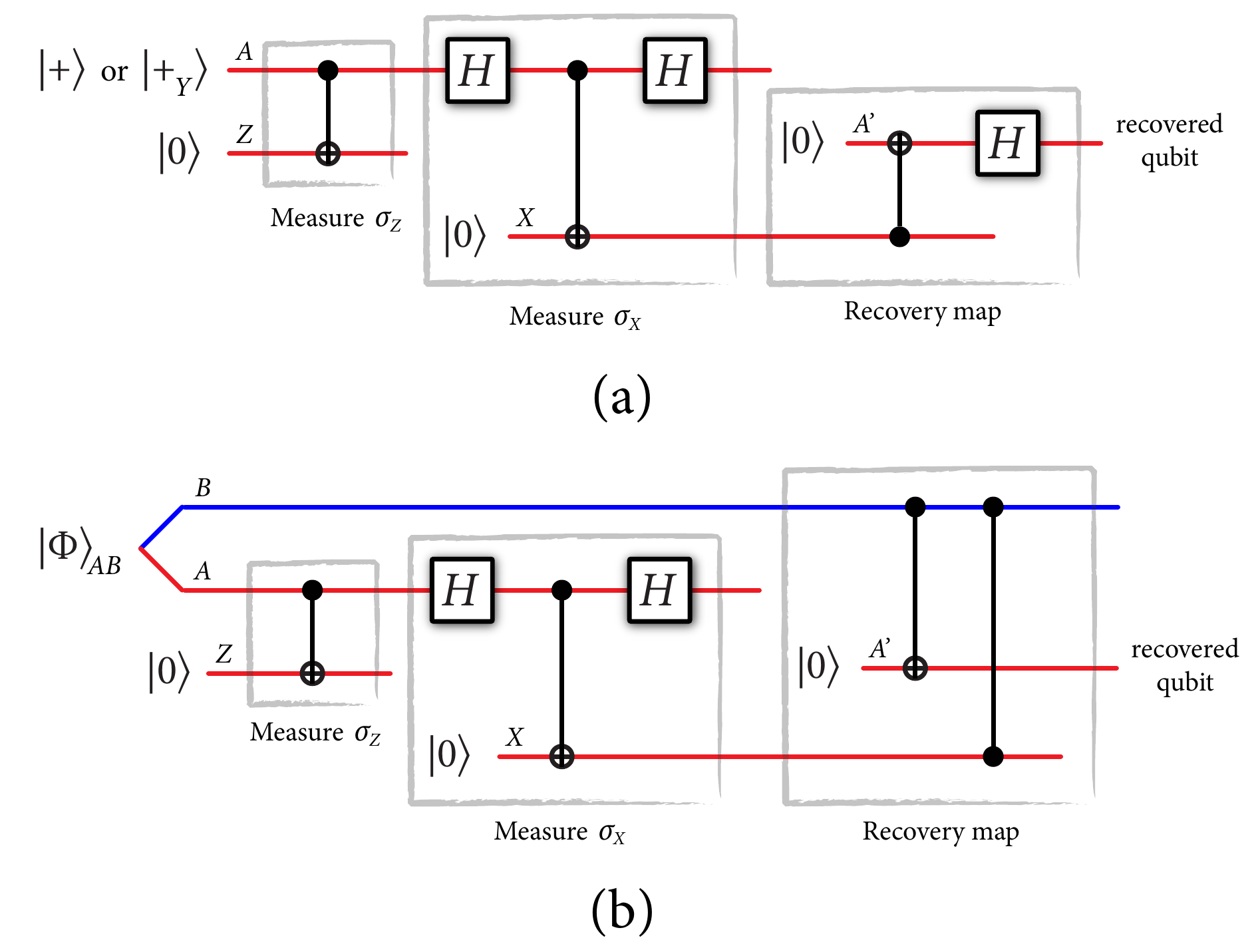}
\end{center}
\caption{Circuits for experimental testing of our entropic uncertainty
relation in \eqref{eq:EUR-QSI-2-new}. \textbf{(a)} Four different experimental
tests in which one can prepare system $A$ as either $|+\rangle_{A}$ or
$|+_{Y}\rangle_{A}$ and then perform either a $\sigma_Z$ Pauli measurement or not.
Afterward, a $\sigma_X$ measurement is performed followed by the recovery operation
from \eqref{eq:r1-recovery-X-eigen}, whose aim it is to undo the effect of the
$\sigma_X$ measurement. If $|+\rangle_{A}$ is prepared and $\sigma_Z$ is not measured, then
it is possible to undo the effect of the $\sigma_X$ measurement and recover the qubit
$|+\rangle$ perfectly in system $A^{\prime}$. If $|+\rangle_{A}$ is prepared
and $\sigma_Z$ is then measured, it is possible to undo the effect of the $\sigma_X$
measurement with the same recovery operation. The same results hold if
$|+_{Y}\rangle_{A}$ is prepared in system $A$ (i.e., the recovery operation
undoes the effect of the $\sigma_X$ measurement). \textbf{(b)} Two different
experimental tests in which one can prepare a maximally entangled Bell state
$|\Phi\rangle_{AB}$ in systems $A$ and $B$, perform a Pauli $\sigma_Z$ measurement or
not, perform a $\sigma_X$ measurement, followed by a recovery operation whose aim it
is to undo the effect of the $\sigma_X$ measurement. In the case that $\sigma_Z$ is not
measured, the recovery operation perfectly restores the maximally entangled
state in systems $A^{\prime}$ and $B$. In the case that $\sigma_Z$ is measured, the
recovery operation undoes the effect of the $\sigma_X$ measurement by restoring the
maximally correlated state $(|00\rangle\langle00|_{A^{\prime}B}+|11\rangle
\langle11|_{A^{\prime}B})/2$ in systems $A^{\prime}$ and $B$. }%
\label{fig:experimental-circuits}%
\end{figure}

We have experimentally tested three of the examples from the previous section,
namely, the $X$ eigenstate, the maximally entangled state, and the $Y$
eigenstate examples. We did so using the recently available IBM \textit{Quantum
Experience} (QE) \cite{IBM16}. Three experiments have already appeared on the arXiv, conducted remotely by theoretical groups testing out experiments which had never been performed previously \cite{AL16,D16,RTSE16}. The QE\ architecture consists of five
fixed-frequency superconducting transmon qubits, laid out in a
\textquotedblleft star geometry\textquotedblright\ (four \textquotedblleft
corner\textquotedblright\ qubits and one in the center). It is possible to
perform single-qubit gates $X$, $Y$, $Z$, $H$, $T$, $S$, and $S^{\dag}$, a
Pauli measurement $Z$, and Bloch sphere tomography on any single qubit.
However, two-qubit operations are limited to controlled-NOT gates with any one
of the corner qubits acting as the source and the center qubit as the target.
Thus, one must \textquotedblleft recompile\textquotedblright\ quantum circuits
in order to meet these constraints. More information about the architecture is
available at the user guide at \cite{IBM16}.

Our experiments realize
and test three of the examples from the previous section and, in particular,
are as follows:

\begin{enumerate}
\item Prepare system $A$ in the state $|+\rangle$. Measure Pauli $\sigma_{X}$
on qubit $A$ and place the outcome in register $X$. Perform the recovery channel
given in \eqref{eq:r1-recovery-X-eigen}, with output system $A^{\prime}$.
Finally, perform Bloch sphere tomography on system $A^{\prime}$.

\item Prepare system $A$ in the state $|+\rangle$. Measure Pauli $\sigma_{Z}$
on qubit $A$ and place the outcome in register $Z$. Measure Pauli $\sigma_{X}$ on
qubit $A$ and place the outcome in register $X$. Perform the recovery channel
given in \eqref{eq:r1-recovery-X-eigen}, with output system $A^{\prime}$.
Finally, perform Bloch sphere tomography on system $A^{\prime}$.

\item Same as Experiment~1 but begin by preparing system $A$ in the state $|+_{Y}%
\rangle_{A}$.

\item Same as Experiment~2 but begin by preparing system $A$ in the state $|+_{Y}%
\rangle_{A}$.

\item Prepare systems $A$ and $B$ in the maximally entangled Bell state
$|\Phi\rangle_{AB}$. Measure Pauli $\sigma_{X}$ on qubit $A$ and place the
outcome in register $X$. Perform the recovery channel given in
\eqref{eq:max-ent-state-recovery-op}, with output systems $A^{\prime}$ and
$B$. Finally, perform measurements of $\sigma_{X}$ on system $A^{\prime}$ and
$\sigma_{X}$ on system $B$, or $\sigma_{Y}$ on system $A^{\prime}$ and
$\sigma_{Y}^{\ast}$ on system $B$, or $\sigma_{Z}$ on system $A^{\prime}$ and
$\sigma_{Z}$ on system $B$.

\item Prepare systems $A$ and $B$ in the maximally entangled Bell state
$|\Phi\rangle_{AB}$. Measure Pauli $\sigma_{Z}$ on qubit $A$ and place the
outcome in register $Z$. Measure Pauli $\sigma_{X}$ on qubit $A$ and place the
outcome in register $X$. Perform the recovery channel given in
\eqref{eq:max-ent-state-recovery-op}, with output systems $A^{\prime}$ and
$B$. Finally, perform measurements of $\sigma_{X}$ on system $A^{\prime}$ and
$\sigma_{X}$ on system $B$, or $\sigma_{Y}$ on system $A^{\prime}$ and
$\sigma_{Y}^{\ast}$ on system $B$, or $\sigma_{Z}$ on system $A^{\prime}$ and
$\sigma_{Z}$ on system $B$.
\end{enumerate}

A quantum circuit that can realize Experiments~1--4 is given in
Figure~\ref{fig:experimental-circuits}(a), and a quantum circuit that can
realize Experiments~5--6 is given in Figure~\ref{fig:experimental-circuits}%
(b). These circuits make use of standard quantum computing gates, detailed in
\cite{book2000mikeandike}, and one can readily verify that they ideally have
the correct behavior, consistent with that discussed for the examples in the
previous section. As stated above, it is necessary to recompile these circuits
into a form which meets the constraints of the QE\ architecture.

\begin{figure}[ptb]
\begin{center}
\includegraphics[
natheight=7.573200in,
natwidth=11.067000in,
height=4.0171in,
width=5.8591in
]{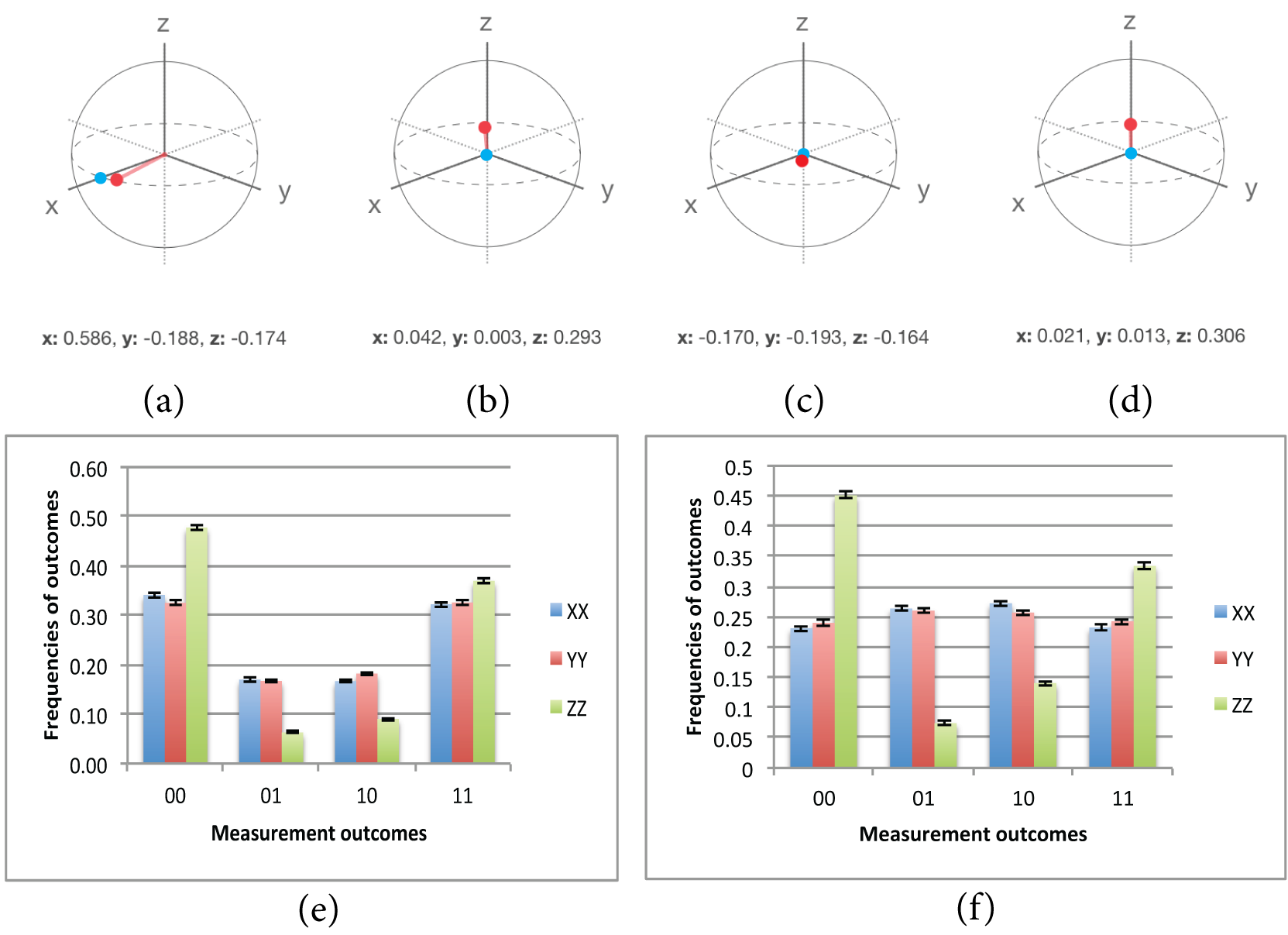}
\end{center}
\caption{Results of experimental tests on the IBM\ QE\ quantum computer.
Subfigures (a)--(f) correspond to Experiments~1--6 outlined in the main text,
respectively. \textbf{(a)} The ideal state of the recovered qubit is
$|+\rangle$, as predicted by \eqref{eq:r1-recovery}
and depicted on the Bloch sphere as a blue dot. The figure plots the result of Bloch sphere tomography from the experimental tests (as a red dot). \textbf{(b)--(d)} The
ideal state of the recovered qubit in each case is $\pi$ (the maximally mixed
state), as predicted by \eqref{eq:r1-recovery-theta-omega},
\eqref{eq:r4-recovery-sigma-rho}, and \eqref{eq:r4-recovery-theta-omega},
respectively (again depicted as blue dots).
The figure again plots the result of Bloch sphere tomography as red dots.
\textbf{(e)}\ The ideal state of the recovered qubits is
$|\Phi\rangle$ as predicted by \eqref{eq:r3-recovery}. In such a case,
measurement of the Pauli observables $\sigma_{i}$ on $A^{\prime}$ and
$\sigma_{i}^{\ast}$ on $B$ for $i\in\{X,Y,Z\}$ should return 00 and 11 with
probability 0.5 and 01 and 10 with probability zero. The plots reveal
significant noise in the experiments, especially from the $\sigma_X$ and $\sigma_Y$ measurements. \textbf{(f)}\ The ideal state of the
recovered qubits is the maximally correlated state $(|00\rangle\langle
00|+|11\rangle\langle11|)/2$ as predicted in
Section~\ref{sec:max-entangled-example}. In such a case, measurement of the
Pauli observables $\sigma_{Z}$ on $A$ and $\sigma_{Z}$ on $B$ should return 00
and 11 with probability 0.5 and 01 and 10 with probability zero. Measurement
of the Pauli observables $\sigma_{i}$ on $A$ and $\sigma_{i}^{\ast}$ on $B$
for $i\in\{X,Y\}$ should return all outcomes with equal probabilities. Again,
the plots reveal significant noise in the experiments.}%
\label{fig:exp-results}%
\end{figure}

Figure~\ref{fig:exp-results}\ plots the results of Experiments 1--6. Each
experiment consists of three measurements, with Experiments 1--4 having
measurements of each of the Pauli operators, and Experiments 5--6 having three
different measurements each as outlined above. Each of these is repeated 8192
times, for a total of $6\times3\times8192=147,456$ experiments. The standard
error for each kind of experiment is thus $\sqrt{p_{c}(1-p_{c})/8192}$, where $p_{c}$
is the estimate of the probability of a given measurement outcome in a given experiment. The caption of
Figure~\ref{fig:exp-results}\ features discussions of and comparisons between
the predictions of the previous section and the experimental outcomes. While
it is clear that the QE\ chip is subject to significant noise, there is still
reasonable agreement with the theoretical predictions of the previous section. One observation we make regarding Figure~\ref{fig:exp-results}(e) is that the frequencies for the outcomes of the $\sigma_Z$ and $\sigma_Z$ measurements are much closer to the theoretically predicted values than are the other measurement outcomes.

\section{Conclusion}

The entropic uncertainty relation with quantum side information is a unifying
principle relating quantum uncertainty due to measurement incompatibility and
entanglement. Here we refine and tighten this inequality with a
state-dependent term related to how well one can reverse the action of a
measurement. The tightening of the inequality is most pronounced when the
measurements and state are all chosen from mutually unbiased bases, i.e., in
our \textquotedblleft maximum uncertainty\textquotedblright\ example with the
measurements being $\sigma_{X}$ and $\sigma_{Z}$ and the initial state being a
$\sigma_{Y}$ eigenstate. We have experimentally tested our theoretical
predictions on the IBM\ \textit{Quantum Experience} and find reasonable
agreement between our predictions and experimental outcomes.

We note that in terms of the conditional min-entropy, other refinements
of~\eqref{eq:EUR-QSI-2} are known~\cite{DFW} that look at the measurement
channel and its own inverse channel, and it would be interesting to understand
their relation. Going forward, it would furthermore be interesting to
generalize the results established here to infinite-dimensional and multiple
measurement scenarios.

\textit{Acknowledgments}---The authors acknowledge discussions with Siddhartha
Das, Michael Walter, and Andreas Winter.
We are grateful to the team at IBM and the IBM Quantum Experience project. This work does not reflect the views or opinions of IBM or any of its employees.
MB acknowledges funding provided by
the Institute for Quantum Information and Matter, an NSF Physics Frontiers
Center (NSF Grant PHY-1125565) with support of the Gordon and Betty Moore
Foundation (GBMF-12500028). Additional funding support was provided by the ARO
grant for Research on Quantum Algorithms at the IQIM (W911NF-12-1-0521). SW
acknowledges support from STW, Netherlands and an NWO VIDI Grant. MMW is
grateful to SW and her group for hospitality during a research visit to QuTech
in May 2015 and acknowledges support from startup funds from the Department of
Physics and Astronomy at LSU, the NSF\ under Award No.~CCF-1350397, and the
DARPA Quiness Program through US Army Research Office award W31P4Q-12-1-0019.

\appendix

\section{Proof of \eqref{eq:EUR-QSI-1-new} and \eqref{eq:EUR-QSI-2-new}}

\label{app:proof}

The main idea of the proof of \eqref{eq:EUR-QSI-1-new} follows the approach
first put forward in \cite{CYZ11} (see also \cite{CCYZ12}), for which the core
argument is the non-increase of quantum relative entropy. Here we instead
apply a refinement of this entropy inequality from \cite{Junge15} (see also
\cite{W15,Sutter15,SBT16}). In order to prove \eqref{eq:EUR-QSI-1-new}, we start by
noting that it suffices to prove it when $\rho_{ABE}=|\psi\rangle\langle
\psi|_{ABE}$ (i.e., the shared state is pure). This is because the conditional
entropy only increases under the discarding of one part of the conditioning
system. We consider the following isometric extensions of the measurement
channels~\cite{stinespring54}, which produce the measurement outcomes and
post-measurement states:
\begin{align}
U_{A\rightarrow XX^{\prime}A}  &  \equiv\sum_{x}|x\rangle_{X}\otimes
|x\rangle_{X^{\prime}}\otimes P_{A}^{x},\\
V_{A\rightarrow ZZ^{\prime}A}  &  \equiv\sum_{z}|z\rangle_{Z}\otimes
|z\rangle_{Z^{\prime}}\otimes Q_{A}^{z}.
\end{align}
We also define the following pure states, which represent purifications of the
states $\sigma_{XBE}$ and $\omega_{ZBE}$\ defined in \eqref{eq:sig_X_state}
and \eqref{eq:om_Z_state}, respectively:%
\begin{align}
|\sigma\rangle_{XX^{\prime}ABE}  &  \equiv U_{A\rightarrow XX^{\prime}A}%
|\psi\rangle_{ABE},\\
|\omega\rangle_{ZZ^{\prime}ABE}  &  \equiv V_{A\rightarrow ZZ^{\prime}A}%
|\psi\rangle_{ABE}.
\end{align}
Consider from duality of conditional entropy for pure states (see, e.g.,
\cite{CCYZ12}) that
\begin{equation}
H(Z|E)_{\omega}=-H(Z|Z^{\prime}AB)_{\omega}=D(\omega_{ZZ^{\prime}AB}\Vert
I_{Z}\otimes\omega_{Z^{\prime}AB}),
\end{equation}
where $D(\rho\Vert\sigma)\equiv\operatorname{Tr}\{\rho\lbrack\log\rho
-\log\sigma]\}$ is the quantum relative entropy \cite{U62}, defined as such
when $\operatorname{supp}(\rho)\subseteq\operatorname{supp}(\sigma)$ and as
$+\infty$ otherwise. Now consider the following quantum channel
\begin{equation}
\mathcal{P}_{ZZ^{\prime}A}(\cdot)\rightarrow\Pi(\cdot)\Pi+(I-\Pi)(\cdot
)(I-\Pi),
\end{equation}
where $\Pi\equiv VV^{\dag}$. From the monotonicity of quantum relative entropy
with respect to quantum channels \cite{Lindblad1975,U77}, we find that
\begin{equation}
D(\omega_{ZZ^{\prime}AB}\Vert I_{Z}\otimes\omega_{Z^{\prime}AB})\geq
D(\mathcal{P}_{ZZ^{\prime}A}(\omega_{ZZ^{\prime}AB})\Vert\mathcal{P}%
_{ZZ^{\prime}A}(I_{Z}\otimes\omega_{Z^{\prime}AB})).
\end{equation}
Consider that $\mathcal{P}_{ZZ^{\prime}A}(\omega_{ZZ^{\prime}AB}%
)=\omega_{ZZ^{\prime}AB}$. Due to the fact that
\begin{equation}
(I-\Pi)\omega_{ZZ^{\prime}AB}(I-\Pi)=0,
\end{equation}
and from the direct sum property of the quantum relative entropy (see, e.g.,
\cite{CCYZ12}), we have that
\begin{equation}
D(\mathcal{P}_{ZZ^{\prime}A}(\omega_{ZZ^{\prime}AB})\Vert\mathcal{P}%
_{ZZ^{\prime}A}(I_{Z}\otimes\omega_{Z^{\prime}AB}))=D(\omega_{ZZ^{\prime}%
AB}\Vert\Pi(I_{Z}\otimes\omega_{Z^{\prime}AB})\Pi).
\end{equation}
Consider that
\begin{equation}
\Pi(I_{Z}\otimes\omega_{Z^{\prime}AB})\Pi=VV^{\dag}(I_{Z}\otimes
\omega_{Z^{\prime}AB})VV^{\dag}=V\!\left(  \sum_{z}Q_{A}^{z}\rho_{AB}Q_{A}%
^{z}\right)  V^{\dag}.
\end{equation}
This, combined with $\omega_{ZZ^{\prime}AB}=V\rho_{AB}V^{\dag}$,\ then implies
that
\begin{align}
D(\omega_{ZZ^{\prime}AB}\Vert\Pi(I_{Z}\otimes\omega_{Z^{\prime}AB})\Pi)  &
=D\!\left(  V\rho_{AB}V^{\dag}\middle\Vert V\!\left(  \sum_{z}Q_{A}^{z}%
\rho_{AB}Q_{A}^{z}\right)  V^{\dag}\right) \\
&  =D\!\left(  \rho_{AB}\middle\Vert\sum_{z}Q_{A}^{z}\rho_{AB}Q_{A}%
^{z}\right)  ,
\end{align}
where the last equality follows from the invariance of quantum relative
entropy with respect to isometries. Now consider the following quantum
channel:
\begin{equation}
\mathcal{M}_{A\rightarrow X}\equiv\operatorname{Tr}_{X^{\prime}A}%
\circ\ \mathcal{U}_{A\rightarrow XX^{\prime}A},
\end{equation}
where $\mathcal{U}_{A\rightarrow XX^{\prime}A}(\cdot)\equiv U(\cdot)U^{\dag}$.
Consider that $\mathcal{M}_{A\rightarrow X}(\rho_{AB})=\sigma_{XB}$. Also, we
can calculate
\begin{equation}
\mathcal{M}_{A\rightarrow X}\!\left(  \sum_{z}Q_{A}^{z}\rho_{AB}Q_{A}%
^{z}\right)
\end{equation}
as follows:
\begin{equation}
(\operatorname{Tr}_{X^{\prime}A}\circ\mathcal{U}_{A\rightarrow XX^{\prime}%
A})\left(  \sum_{z}Q_{A}^{z}\rho_{AB}Q_{A}^{z}\right)  =\theta_{XB}.
\end{equation}
From~\cite{Junge15}, we have the following inequality holding for a density
operator $\rho$, a positive semi-definite operator $\sigma$, and a quantum
channel $\mathcal{N}$:
\begin{equation}
D(\rho\Vert\sigma)-D(\mathcal{N}(\rho)\Vert\mathcal{N}(\sigma))\geq-\log
F(\rho,\mathcal{R}(\mathcal{N}(\rho))), \label{eq:mono-RE-refine}%
\end{equation}
where $\operatorname{supp}(\rho)\subseteq\operatorname{supp}(\sigma)$ and
$\mathcal{R}$ is a recovery channel with the property that $\mathcal{R}%
(\mathcal{N}(\sigma))=\sigma$. Specifically, $\mathcal{R}$ is what is known as
a variant of the Petz recovery channel, having the form
\begin{align}
\mathcal{R}(\cdot)\equiv &  \int\mathrm{d}t\,p(t)\,\sigma^{-it/2}%
\mathcal{R}_{\sigma,\mathcal{N}}(\mathcal{N}(\sigma)^{it/2}(\cdot
)\mathcal{N}(\sigma)^{-it/2})\sigma^{it/2}\nonumber\\
&  \mathrm{with}\quad p(t)\equiv\frac{\pi}{2}(\cosh(\pi t)+1)^{-1},
\label{eq:petz-recovery-rotated}%
\end{align}
where $\mathcal{R}_{\sigma,\mathcal{N}}$ is the Petz recovery channel
\cite{Petz1986,Petz1988,HJPW04}\ defined as
\begin{equation}
\mathcal{R}_{\sigma,\mathcal{N}}(\cdot)\equiv\sigma^{1/2}\mathcal{N}^{\dag
}\!\left(  \mathcal{N}(\sigma)^{-1/2}(\cdot)\mathcal{N}(\sigma)^{-1/2}\right)
\sigma^{1/2},
\end{equation}
with $\mathcal{N}^{\dag}$ the adjoint of $\mathcal{N}$ (with respect to the
Hilbert--Schmidt inner product). Applying this to our case, we find that
\begin{multline}
D\!\left(  \rho_{AB}\middle\Vert\sum_{z}Q_{A}^{z}\rho_{AB}Q_{A}^{z}\right)
\geq D\!\left(  \mathcal{M}_{A\rightarrow X}(\rho_{AB})\middle\Vert
\mathcal{M}_{A\rightarrow X}\!\left(  \sum_{z}Q_{A}^{z}\rho_{AB}Q_{A}%
^{z}\right)  \right) \label{eq:EUR-QSI-mono-RE-refine}\\
-\log F(\rho_{AB},\mathcal{R}_{XB\to AB}(\mathcal{M}_{A\rightarrow X}%
(\rho_{AB}))),
\end{multline}
where the recovery channel is such that%
\begin{equation}
\mathcal{R}_{XB\to AB}\!\left(  \mathcal{M}_{A\rightarrow X}\!\left(  \sum
_{z}Q_{A}^{z}\rho_{AB}Q_{A}^{z}\right)  \right)  =\sum_{z}Q_{A}^{z}\rho
_{AB}Q_{A}^{z}. \label{eq:Z-X-recover-perfect-general}%
\end{equation}
Consider from our development above that
\begin{align}
D\!\left(  \mathcal{M}_{A\rightarrow X}(\rho_{AB})\middle\Vert\mathcal{M}%
_{A\rightarrow X}\!\left(  \sum_{z}Q_{A}^{z}\rho_{AB}Q_{A}^{z}\right)
\right)   &  =D(\sigma_{XB}\Vert\theta_{XB})\\
&  \geq D(\sigma_{XB}\Vert I_{X}\otimes\sigma_{B})-\log c,
\end{align}
where we have used $\sigma\leq\sigma^{\prime}\Rightarrow D(\rho\Vert
\sigma^{\prime})\leq D(\rho\Vert\sigma)$ (see, e.g., \cite{CCYZ12}), applied
to $Q_{A}^{z}P_{A}^{x}Q_{A}^{z}=|Q_{A}^{z}P_{A}^{x}|^{2}\leq c\cdot I_{A}$,
with $c$ defined in~\eqref{eq:c-param}. Putting everything together, we
conclude that%
\begin{equation}
D(\omega_{ZZ^{\prime}AB}\Vert I_{Z}\otimes\omega_{Z^{\prime}AB})\geq
D(\sigma_{XB}\Vert I_{X}\otimes\sigma_{B})-\log c-\log F(\rho_{AB}%
,\mathcal{R}_{XB\to AB}(\sigma_{XB})),
\end{equation}
which, after a rewriting, is equivalent to \eqref{eq:EUR-QSI-1-new} coupled
with the constraint in \eqref{eq:Z-X-recover-perfect-general}.

The inequality in \eqref{eq:EUR-QSI-2-new} follows from
\eqref{eq:EUR-QSI-1-new} by letting $|\psi\rangle_{ABE}$ be a purification of
$\rho_{AB}$ and observing that
\begin{equation}
H(Z|E)_{\omega}-H(Z|B)_{\omega}=-H(A|B)_{\rho},
\end{equation}
whenever $\rho_{ABE}$ is a pure state and $Q_{A}^{z}=|z\rangle\langle z|_{A}$
for some orthonormal basis $\{|z\rangle_{A}\}_{z}$.

\section{Explicit form of recovery map}

\label{sec:recovery-map-explicit}Here we establish the explicit form given in
\eqref{eq:main:recovery1} for the recovery map, in the case that $\{Q_{A}%
^{z}=|z\rangle\langle z|_{A}\}$ for some orthonormal basis $\{|z\rangle
_{A}\}_{z}$. The main idea is to determine what $\mathcal{R}_{XB\to AB}$ in
\eqref{eq:EUR-QSI-mono-RE-refine} should be by inspecting
\eqref{eq:mono-RE-refine} and \eqref{eq:petz-recovery-rotated}. For our setup,
we are considering a bipartite state $\rho_{AB}$, a set $\{Q_{A}^{z}\}$\ of
measurement operators, and the measurement channel%
\begin{equation}
\mathcal{M}_{A\rightarrow X}(\zeta_{A})\equiv\sum_{x}\operatorname{Tr}%
\{P_{A}^{x}\zeta_{A}\}|x\rangle\langle x|_{X},
\end{equation}
where $\{P_{A}^{x}\}_{x}$ is a set of projective measurement operators. The
entropy inequality in \eqref{eq:EUR-QSI-mono-RE-refine}\ reduces to%
\begin{multline}
D\!\left(  \rho_{AB}\middle\Vert\sum_{z}|z\rangle\langle z|_{A}\otimes
\omega_{B}^{z}\right)  -D\!\left(  \mathcal{M}_{A\rightarrow X}(\rho
_{AB})\middle\Vert\sum_{x}|x\rangle\langle x|_{X}\otimes\theta_{B}^{x}\right)
\\
\geq-\log F(\rho_{AB},\mathcal{R}_{XB\rightarrow AB}(\mathcal{M}_{A\rightarrow
X}(\rho_{AB}))),
\end{multline}
where%
\begin{equation}
\omega_{B}^{z}\equiv(\langle z|_{A} \otimes I_{B})\rho_{AB}(|z\rangle_{A}
\otimes I_{B}),\qquad\theta_{B}^{x}\equiv\sum_{z}\langle z|_{A}P_{A}%
^{x}|z\rangle_{A}\omega_{B}^{z}.
\end{equation}
Observe that%
\begin{equation}
\sum_{x}|x\rangle\langle x|_{X}\otimes\theta_{B}^{x}=\mathcal{M}_{A\rightarrow
X}\!\left(  \sum_{z}|z\rangle\langle z|_{A}\otimes\omega_{B}^{z}\right)  .
\end{equation}
Writing the measurement channel as%
\begin{align}
\mathcal{M}_{A\rightarrow X}(\zeta_{A})  &  \equiv\sum_{x}\operatorname{Tr}%
\{P_{A}^{x}\zeta_{A}P_{A}^{x}\}|x\rangle\langle x|_{X}=\sum_{x,j}\langle
j|_{A}P_{A}^{x}\zeta_{A}P_{A}^{x}|j\rangle_{A}|x\rangle\langle x|_{X}\\
&  =\sum_{x,j}|x\rangle_{X}\langle j|_{A}P_{A}^{x}\zeta_{A}P_{A}^{x}%
|j\rangle_{A}\langle x|_{X},
\end{align}
we can see that a set of Kraus operators for it is $\{|x\rangle_{X}\langle
j|_{A}P_{A}^{x}\}_{x,j}$. So its adjoint is as follows:%
\begin{align}
\left(  \mathcal{M}_{A\rightarrow X}\right)  ^{\dag}(\kappa_{X})  &
=\sum_{x,j}P_{A}^{x}|j\rangle_{A}\langle x|_{X}\kappa_{X}|x\rangle_{X}\langle
j|_{A}P_{A}^{x}=\sum_{x,j}\langle x|_{X}\kappa_{X}|x\rangle_{X}P_{A}%
^{x}|j\rangle_{A}\langle j|_{A}P_{A}^{x}\\
&  =\sum_{x}\langle x|_{X}\kappa_{X}|x\rangle_{X}P_{A}^{x}.
\end{align}
So by inspecting \eqref{eq:mono-RE-refine} and
\eqref{eq:petz-recovery-rotated}, we see that the recovery map has the
following form:%
\begin{align}
&  \mathcal{R}_{XB\rightarrow AB}(\xi_{XB})\nonumber\\
&  =\int dt\ p(t)\ \left(  \sum_{z}|z\rangle\langle z|_{A}\otimes\omega
_{B}^{z}\right)  ^{\frac{1-it}{2}}\sum_{x}P_{A}^{x}\left(  \langle
x|_{X}\otimes I_{B}\right)  \left(  \sum_{x^{\prime}}|x^{\prime}\rangle\langle
x^{\prime}|_{X}\otimes\theta_{B}^{x^{\prime}}\right)  ^{\frac{-1+it}{2}}%
(\xi_{XB})\nonumber\\
&  \qquad\qquad\left(  \sum_{x^{\prime\prime}}|x^{\prime\prime}\rangle\langle
x^{\prime\prime}|_{X}\otimes\theta_{B}^{x^{\prime\prime}}\right)
^{\frac{-1-it}{2}}\left(  |x\rangle_{X}\otimes I_{B}\right)  \left(
\sum_{z^{\prime}}|z^{\prime}\rangle\langle z^{\prime}|_{A}\otimes\omega
_{B}^{z^{\prime}}\right)  ^{\frac{1+it}{2}}\\
&  =\int dt\ p(t)\ \left(  \sum_{z}|z\rangle\langle z|_{A}\otimes\left(
\omega_{B}^{z}\right)  ^{\frac{1-it}{2}}\right)  \sum_{x}P_{A}^{x}\left(
\langle x|_{X}\otimes I_{B}\right) \nonumber\\
&  \qquad\qquad\left[  \left(  \sum_{x^{\prime}}|x^{\prime}\rangle\langle
x^{\prime}|_{X}\otimes\left(  \theta_{B}^{x^{\prime}}\right)  ^{\frac
{-1+it}{2}}\right)  (\xi_{XB})\left(  \sum_{x^{\prime\prime}}|x^{\prime\prime
}\rangle\langle x^{\prime\prime}|_{X}\otimes\left(  \theta_{B}^{x^{\prime
\prime}}\right)  ^{\frac{-1-it}{2}}\right)  \right] \nonumber\\
&  \qquad\qquad\left(  |x\rangle_{X}\otimes I_{B}\right)  \left(
\sum_{z^{\prime}}|z^{\prime}\rangle\langle z^{\prime}|_{A}\otimes\left(
\omega_{B}^{z^{\prime}}\right)  ^{\frac{1+it}{2}}\right)
\end{align}%
\begin{align}
&  =\int dt\ p(t)\ \left(  \sum_{z}|z\rangle\langle z|_{A}\otimes\left(
\omega_{B}^{z}\right)  ^{\frac{1-it}{2}}\right)  \sum_{x,x^{\prime}%
,x^{\prime\prime}}P_{A}^{x}\langle x|_{X}|x^{\prime}\rangle\langle x^{\prime
}|_{X}\otimes\left(  \theta_{B}^{x^{\prime}}\right)  ^{\frac{-1+it}{2}}%
(\xi_{XB})\nonumber\\
&  \qquad\qquad|x^{\prime\prime}\rangle\langle x^{\prime\prime}|_{X}%
|x\rangle_{X}\otimes\left(  \theta_{B}^{x^{\prime\prime}}\right)
^{\frac{-1-it}{2}}\left(  \sum_{z^{\prime}}|z^{\prime}\rangle\langle
z^{\prime}|_{A}\otimes\left(  \omega_{B}^{z^{\prime}}\right)  ^{\frac{1+it}%
{2}}\right) \\
&  =\int dt\ p(t)\ \left(  \sum_{z}|z\rangle\langle z|_{A}\otimes\left(
\omega_{B}^{z}\right)  ^{\frac{1-it}{2}}\right)  \sum_{x}P_{A}^{x}\langle
x|_{X}\otimes\left(  \theta_{B}^{x}\right)  ^{\frac{-1+it}{2}}(\xi
_{XB})|x\rangle_{X}\otimes\left(  \theta_{B}^{x}\right)  ^{\frac{-1-it}{2}%
}\nonumber\\
&  \qquad\qquad\left(  \sum_{z^{\prime}}|z^{\prime}\rangle\langle z^{\prime
}|_{A}\otimes\left(  \omega_{B}^{z^{\prime}}\right)  ^{\frac{1+it}{2}}\right)
\\
&  =\int dt\ p(t)\ \sum_{z,x,z^{\prime}}|z\rangle\langle z|_{A}P_{A}%
^{x}|z^{\prime}\rangle\langle z^{\prime}|_{A}\otimes\left(  \omega_{B}%
^{z}\right)  ^{\frac{1-it}{2}}\left(  \theta_{B}^{x}\right)  ^{\frac{-1+it}%
{2}}\operatorname{Tr}_{X}\{|x\rangle\langle x|_{X}(\xi_{XB})\}\left(
\theta_{B}^{x}\right)  ^{\frac{-1-it}{2}}\left(  \omega_{B}^{z^{\prime}%
}\right)  ^{\frac{1+it}{2}}.
\end{align}
We can thus abbreviate its action as%
\begin{equation}
\mathcal{R}_{XB\rightarrow AB}(\xi_{XB})=\ \sum_{z,x,z^{\prime}}%
|z\rangle\langle z|_{A}P_{A}^{x}|z^{\prime}\rangle\langle z^{\prime}%
|_{A}\otimes\mathcal{R}_{XB\rightarrow B}^{x,z,z^{\prime}}(\xi_{XB}),
\end{equation}
where%
\begin{equation}
\mathcal{R}_{XB\rightarrow B}^{x,z,z^{\prime}}(\xi_{XB})\equiv\int
dt\ p(t)\left(  \omega_{B}^{z}\right)  ^{\frac{1-it}{2}}\left(  \theta_{B}%
^{x}\right)  ^{\frac{-1+it}{2}}\operatorname{Tr}_{X}\{|x\rangle\langle
x|_{X}(\xi_{XB})\}\left(  \theta_{B}^{x}\right)  ^{\frac{-1-it}{2}}\left(
\omega_{B}^{z^{\prime}}\right)  ^{\frac{1+it}{2}}.
\end{equation}
(Note that $\mathcal{R}_{XB\rightarrow B}^{x,z,z^{\prime}}$ is not a channel.)
So then the action on the classical--quantum state $\sigma_{XB}$, defined as%
\begin{equation}
\sigma_{XB}\equiv\sum_{x}|x\rangle\langle x|_{X}\otimes\sigma_{B}^{x},
\end{equation}
with $\sigma_{B}^{x}\equiv\operatorname{Tr}_{A}\{P_{A}^{x}\rho_{AB}\}$, is as
follows:%
\begin{equation}
\mathcal{R}_{XB\rightarrow AB}(\sigma_{XB})=\sum_{z,x,z^{\prime}}%
|z\rangle\langle z|_{A}P_{A}^{x}|z^{\prime}\rangle\langle z^{\prime}%
|_{A}\otimes\int dt\ p(t)\left(  \omega_{B}^{z}\right)  ^{\frac{1-it}{2}%
}\left(  \theta_{B}^{x}\right)  ^{\frac{-1+it}{2}}\sigma_{B}^{x}\left(
\theta_{B}^{x}\right)  ^{\frac{-1-it}{2}}(\omega_{B}^{z^{\prime}}%
)^{\frac{1+it}{2}}.
\end{equation}

\bibliographystyle{unsrt}
\bibliography{Ref}

\end{document}